# A Milli-Newton Propulsion System for the Asteroid Mobile Imager and Geologic Observer (AMIGO)

Greg Wilburn
Space and Terrestrial Robotic
Exploration (SpaceTREx) Laboratory
Aerospace and Mechanical
Engineering Dept.
University of Arizona
gregwilburn@email.arizona.edu

Erik Asphaug
Lunar and Planetary Laboratory
College of Science
University of Arizona
asphaug@lpl.arizona.edu

Jekan Thangavelautham
Space and Terrestrial Robotic
Exploration (SpaceTREx) Laboratory
Aerospace and Mechanical
Engineering Dept.
University of Arizona
jekan@email.arizona.edu

*Abstract*— Exploration of small bodies, namely comets and asteroids remain a challenging endeavor due to their low gravity. The risk is so high that missions such as Hayabusa II and OSIRIS-REx will be performing touch and go missions to obtain samples. The next logical step is to perform longer-term mobility on the surface of these asteroid. This can be accomplished by sending small landers of a 1 kg or less with miniature propulsion systems that can just offset the force of asteroid gravity. Such a propulsion system would ideally be used to hop on the surface of the asteroid. Hopping has been found to be most efficient form of mobility on low-gravity. Use of wheels for rolling presents substantial challenges as the wheel can't gain traction to roll. The Asteroid Mobile Imager and Geologic Observer (AMIGO) utilizes 1 kg landers that are stowed in a 1U CubeSat configuration and deployed, releasing an inflatable that is 1-m in diameter. The inflatable is attached to the top of the 1U lander, enabling high speed communications and a means of easily tracking lander from an overhead mothership. Milligravity propulsion is required for the AMIGO landers to perform ballistic hops on the asteroid surface. The propulsion system is used to navigate the lander across the surface of the asteroid under the extremely low gravity while taking care to not exceed escape velocity. Although the concept for AMIGO missions is to use multiple landers, the more surface area evaluated by each lander the better. Without a propulsion system, each AMIGO will have a limited range of observable area. The propulsion system also serves as a rough attitude control system (ACS), as it enables pointing and regulation over where the lander is positioned via an array of MEMS thrusters. Several different techniques have been proposed for hopping nano-landers on low gravity environments including use of reaction wheels, electro-polymers, and rocket thrusters. In this concept, we will be heating sublimate to provide propulsive thrust which is simple and effective. Storing the propulsive gas as a solid provides much better storage density to ensure the longest lifetime possible. The starting point for this propulsion system involves selection of an appropriate, high-performance sublimate that meets the mobility needs of AMIGO. The paper will cover selection of the right sublimate, initial design and prototyping of the thruster using standard off-the-shelf arrays of electro-active MEMs valves. The device will be integrated with the fuel source with thrust profiles measured inside a milli-newton test stand inside a vacuum chamber. MEMS devices are to be used because of the weight and volume saving potential compared to traditional large-sized thrusting mechanisms. Through these efforts we are advancing on a milli-newton thruster that use non-combustibles and can be readily integrated onto nano-landers for asteroid surface exploration.

TABLE OF CONTENTS

## 1. INTRODUCTION

On the surface of an asteroid, local gravity is low enough that a traditional rover's wheels rotating would cause a delta-v high enough to reach escape velocity of the asteroid, not to mention that the wheels will not have traction on the loosely cohesive surface. As such, asteroid landers often must hop across the surface to achieve mobility.

JAXA's MINERVA-II 1A and 1B landers on the Hayabusa 2 mission to asteroid Ryugu utilize reaction wheels to convert angular momentum into a horizontal propulsive force on the lander. Figure 1 shows an image by MINERVA-II 1B shortly after landing on Ryugu's surface.

The delta-v to reach escape velocity from an approximate radius of 400m and mass $4.5 \times 10^{11}$ kg is 0.017 m/s. Some factor of safety is applied to ensure escape velocity is not reached. The dynamics of the surface regolith must be well understood for accurate prediction of hopping dynamics, as the force transferred to the robot is dependent on robot-regolith interaction. Some analysis has been done by JPL on their proposed Hedgehog hopper in low gravity simulations on zero-g airplane maneuvers. There are also plans to use a low-speed on-orbit centrifuge to simulate asteroid surface gravity and test mobility platforms [33-35]. It is desired to have the hopping of the spacecraft independent of the surface characteristics, so this momentum transfer mechanism is not considered for further analysis.

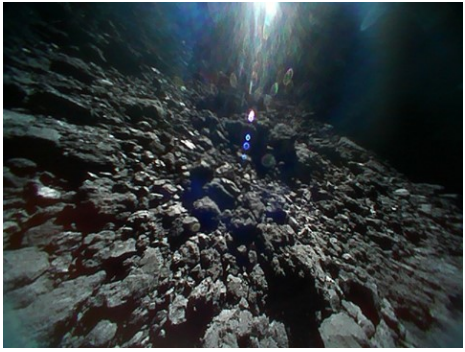

**Figure 1:** Surface of Ryugu from MINERVA-II 1B (Credit: JAXA, University of Tokyo et al.).

Low thrust, low mass propulsion systems are vital to enable exploration of asteroid as conventional chemical rockets thrusters are often too bulky and complex for the limited volume and mass budget of nano-landers. Thus cold gas and MEMS thrusters are explored.

Our own efforts in hopping-robots have resulted in several designs intended for the range of surface gravity environments on small bodies. We are developing the Sphere-X robotic hopper/free-flyer for lunar and Mars extreme environment exploration [21-24]. For the Sphere-X platform we have considered chemical propulsion and mechanical hopping mechanisms [21-22, 23]. For lower gravity asteroid surfaces, we have been developing the nano-lander platform that uses onboard cold-gas or reaction wheels to hop for short duration [25]. AMIGO is advancement over our nano-lander concept, as it contains a multi-functional inflatable [29-32] for use in mobility, communications and tracking and has a propulsion system finely tuned to the asteroid low-gravity environment [26]. AMIGO by hopping and rolling will be able to navigate the maze-like surface pathways using autonomous control [36].

Generally, nozzles can be arranged as an array on a single chip to provide thrust and torque on a spacecraft. Using this discretized nozzle system, rough attitude control is possible with small spacecraft by actuating individual thrusters. Firing all nozzles at once would provide a larger thrust and act as if it were a single nozzle aligned with the center of gravity of a small spacecraft. Table 1 shows a comparison of these previously developed micro-propulsion cold gas and electrothermal systems:

**Table 1: Current Micro-Thrusters**

| System | Type | Propellant | $I_{sp}$ [s] | Thrust [mN] |
|---|---|---|---|---|
| MSPT [1] | Sublimate Resistojet | Lead-Styphnate | NA | 3619 |
| VLM [2] | Liquid Resistojet | Water | 119-124 | 0.68-1 |
| FMMR [3] | Liquid Resistojet | Water | 79 | 1.7 |
| GOMSpace MEMS [4] | Cold Gas | Methane | 50-75 | 1 |
| CNAPS [5] | Liquid Cold Gas | $SF_6$ | 35 | 10-40 |
| SNAP 1 [6] | Liquid Cold Gas | Butane | 43 | 50 |
| CHIPS [7] | Liquid Cold Gas | R-134a | 82 | 30 |
| AMR [8] | Liquid Cold Gas | R-134a | 150 | 10 |
| μPS [9] | Liquid Cold Gas | Water | 100 | .5-3 |

The Micro Solid Propellant Thruster (MSPT) developed by Lee et al. [1] is an array of one time use thrusters. The operating principal is that a solid propellant is heated by a platinum ignition coil to build pressure behind a diaphragm. As the pressure reaches some critical burst pressure of the diaphragm, the diaphragm bursts and releases the gas to provide thrust. Ignition of the sublimate was able to produce a max thrust of 3619 milli-Newtons. Mechanical design is fairly simple as there does not need to be any flow channels or control valves between the storage tank and nozzle due to clever design. The negative of this design is that each nozzle can only be used once, limiting the mission duration. Use of water for propulsion is another viable option [2-3, 27-28]. Water is found in abundance on C-type asteroids, however it will require extra effort to extract and refine it for use as a propellant.

The Free Molecule Micro Resistojet [3] utilizes electrical heating of individual molecules. Instead of a converging-diverging nozzle, propellant passes through an array of electrically heated slots. The slot geometry is smaller than the mean free path of the molecules, heating the molecules when they come in contact with the wall.

An interesting comparison of two systems, CHIPS and AMR, using the same propellant R-134a resulting in differing thrust and specific impulse. CHIPS has a lower specific impulse with higher thrust contrasting to AMR's results. The CHIPS system would be able to lift a heavier robot of the surface of some small solar system body, while AMR can efficiently propel smaller systems. From the Tsiolkovsky rocket equation, AMR can provide more delta-v, thus extending the life of a hopping system. A more suitable solution for hopping on an asteroid must be developed. In the following section we present the AMIGO asteroid lander concept, followed by evaluation of cold-gas and electrothermal propulsion options in Section 3, sublimates in Section 4 followed by geometric approximation of the nozzle in Section 5, simulations in Section 6, discussions in Section 7 and conclusions in Section 8.

## 2. AMIGO-An Asteroid Hopper

The Asteroid Mobile Imager and Geologic Observer (AMIGO) (Figures 3-5) that will be deployed at multiple locations around the surfaces of small bodies and provide stereo imaging from vantage points ~1 m above the surface, close-up geologic imaging, and seismic sensing. The payload consists of three or more 1U CubeSats that each contains an inflatable and on-board propulsion system to perform surface hopping. The concept of operations is shown in Figure 6.

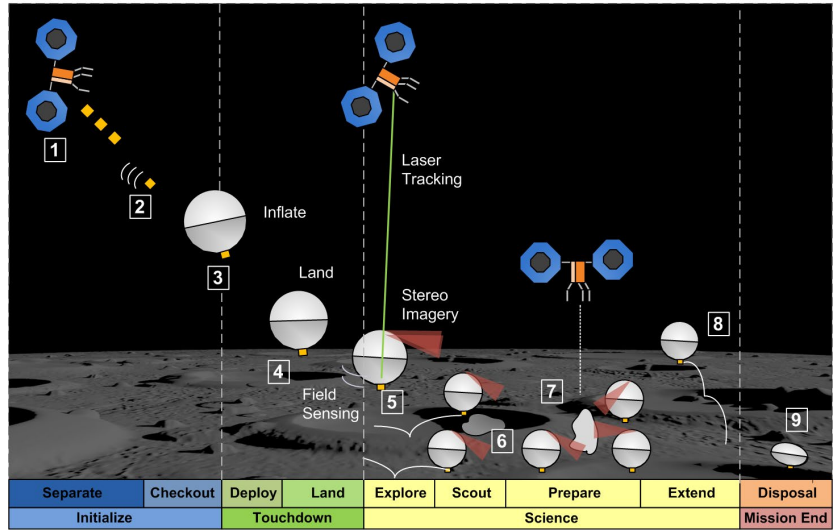

Figure 6: AMIGO Concept of Operations

Each AMIGO is deployed from a mother spacecraft. During descent, the robot inflates from its stowed 1U state. Upon landing, initial context is determined for where the robot is on the surface. This is done by both on board imaging and tracking from the mother. The inflatable portion provides a tracking target, as smaller robots may not be large enough to be tracked. From there, the science mission is conducted. For the AMIGO lander, there are five science goals:

1. Determine local surface hardness and compliance
2. Acquire seismic data constraining the geologic competence of the asteroid
3. Acquire micro-imaging of fine geologic structure from diverse locations
4. Detect images of thermal fatigue of surface rocks
5. Measure electric fields and properties of surface regolith

Each of these science goals seeks to fill a current knowledge gap in the characteristics of asteroids. For example, the proposed NASA Asteroid Redirect Mission was to retrieve a boulder from the surface of a near Earth asteroid and return the sample for further analysis [10]. Currently, the dynamics of how to extract a boulder from the surface of an asteroid is an open problem. The issue is as fundamental as Newton's Third Law; if one aims to pull a three-ton boulder from the asteroid surface, the spacecraft must exert three tons on the asteroid. Will the asteroid and boulder have enough cohesive strength to not completely fall apart? Seismic sensors and close-up geologic sensors will provide this information.

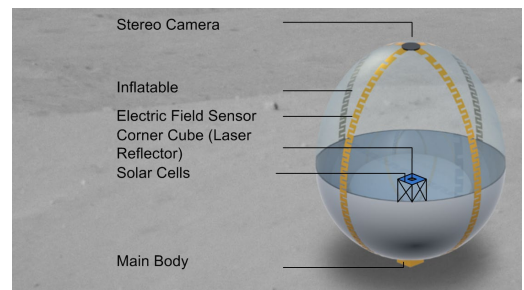

Figure 3: AMIGO Lander

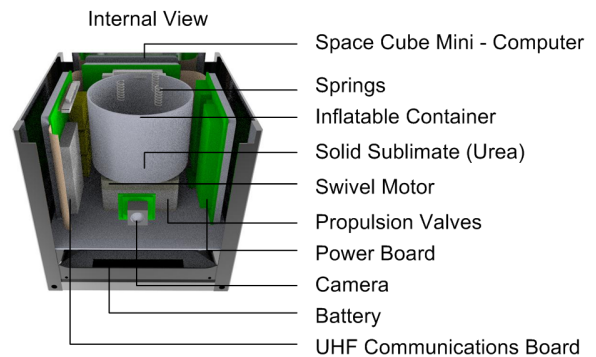

Figure 4: AMIGO Internals

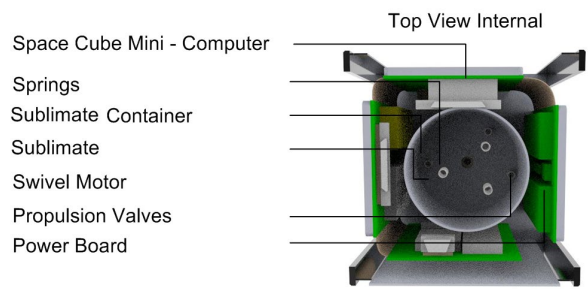

Figure 5: Top View of AMIGO Internals

The characterization of surface regolith of asteroids is vital to the success of future lander missions and the further understanding of the composition of asteroids. For instance, it is theorized that planetesimals often impacted with each other and either obliterated into fine dust and small clumps or aggregated together. In either case, fine grains are created. For intact planetesimals, this dust accreted to the surface and became the surface regolith. However, that regolith may not have the same compositions as the asteroid itself due to being a combination of multiple meteoric impact events. In situ analysis will aid in the understanding of the surface of asteroids in this regard. A large reason for the concept of AMIGO is to add to the current base of knowledge for the surface characteristics of asteroids for use in future lander missions. The familiarity with asteroid surfaces gained by lower cost missions will lay the foundation for, say, a Discovery class mission to be more successful due to limiting the unknowns in the geology dynamics of asteroids.

## 3. COLD GAS / ELECTROTHERMAL PROPULSION

Cold gas systems tend to be the simplest in design amongst propulsion technology. In its simplest form, the only required components are a storage tank, isolation valve, and nozzle. Electrothermal systems, namely resistojets, can be thought of as a cold gas system with heat addition either in the storage tank or flow ducts. Examination if the isentropic relations [11] provides insight into nozzle geometry and propellant selection. Delivered thrust is

$$F = \lambda \dot{m} v_e + (p_e - p_a)A_e \tag{1}$$

Where F is the thrust, $\lambda$ is the conical nozzle correction factor accounting non-axial flow, $\dot{m}$ is the mass flow rate of propellant, $v_e$ is exhaust velocity, $p_e$ is nozzle exit pressure, $p_a$ is ambient pressure, and $A_e$ is the nozzle exit area. $\lambda$ is based on the semi-vertex angle $\sigma$ of the cone nozzle,

$$\lambda = \frac{1 + \cos(\sigma)}{2} \tag{2}$$

The equation for exhaust velocity is:

$$v_{e_i} = \sqrt{\frac{2\gamma R_{univ} T_c}{M(\gamma - 1)}\left(1 - \frac{T_e}{T_c}\right)} \tag{3}$$

The equation for mass flow rate is:

$$\dot{m}_i = A_t p_c \gamma \sqrt{\frac{\left(\frac{2}{\gamma + 1}\right)^{\frac{\gamma+1}{\gamma-1}}}{\gamma R T_c}} \tag{4}$$

Where $\gamma$ is the specific heat ratio, $R_{univ}$ is the universal gas constant, $T_c$ is the chamber temperature, $T_e$ is the exit temperature, M is the molecular mass of the gas, $A_t$ is the nozzle throat area, $p_c$ is the chamber pressure, and R is the gas constant for a specific gas. The i subscript refers to isentropic assumptions. For cold gas systems, the ideal specific impulse $I_{sp}$ is analyzed to compare efficiency of potential propellants,

$$I_{sp} = \frac{v_e}{g_0} \tag{5}$$

**Table 2: $I_{sp}$ of Cold Gas Propellants**

| Gas | $\gamma$ | M (g/mol) | $I_{sp}$ (s) |
| --- | --- | --- | --- |
| $SF_6$ | 1.11 | 146 | 76 |
| Air | 1.4 | 28.97 | 91 |
| $N_2$ | 1.4 | 28.01 | 85 |
| Ar | 1.66 | 39.95 | 57 |
| $H_2$ | 1.41 | 2 | 296 |
| $CH_4$ | 1.32 | 16 | 114 |
| $C_4H_{10}$ | 1.09 | 58.12 | 80 |
| $I_2$ | 1.26 | 253.8 | 40 |

Except for hydrogen, which is not used in practice due to storage issues, cold gas systems are limited to 115s $I_{sp}$. In actual systems, the limitation is to 80 s due to losses. Resistojets increase $I_{sp}$ by increasing the chamber temperature. This in turn leads to higher exhaust velocities. In general, lower molecular weight and higher specific heat ratio leads to a higher efficiency. Some resistojets can achieve an $I_{sp}$ over 200 s [12].

Conventionally, cold gas propellant is stored as a gas, common with systems utilizing nitrogen or hydrogen. The issue with gas storage is that the storage tank must be at high pressures to contain an appreciable amount of fuel for longer mission durations. Higher pressure systems are prone to leak much more than lower pressure tanks, as leak rate is proportional to the storage pressure. This means components must be bulkier to withstand stress and the increased volume of gas. Two other methods of cold gas storage are liquid phase and solid phase. These two approaches work on the principle of vapor pressure equilibrium; for any given temperature, the gas will exert some pressure on the condensed phase. For liquids, this is referred to as evaporation; solids sublimate into a gas phase. Both storage schemes have the benefit of lower pressure tanks compared to gas storage, though some additional components are required.

The vapor pressure for liquids tends to be higher than solids for a given temperature. Liquid storage has been utilized to great effect, notably with water in various vaporizing liquid microthrusters and sulfur hexafluoride (SF6) for the NANOPS system of the Can-X2 mission [13]. The downside compare to sublimates is the issue of "sloshing," where the liquid freely moves in the storage tank causing inertial instabilities and potential clogging of the tank exit to flow channels. Sublimates are able to be fixed in place to the storage tank and deserve consideration.

## 4. SUBLIMATE PROPELLANTS

For substances below their triple point in a phase diagram, a solid can phase transition to a gas without transitioning to an intermediate liquid phase in a process called sublimation; the opposite process is deposition. For any pressure and temperature the solid is subjected to, there exists an equilibrium vapor pressure, or solid-gas equilibrium pressure, that the gaseous state exerts on the solid. This pressure is not to be confused with the ambient pressure, which is the pressure exerted on the entire two-phase system. The vapor pressure is a nonlinear function of temperature, and can be approximated by the Clausius-Clapeyron Relation:

$$\frac{dP}{dT} = \frac{L}{T \Delta v} \quad (6)$$

Where P is the pressure, T is temperature, L is specific latent heat, and v is specific volume. There are multiple derived forms of this equation, such as the Antoine Equation whose coefficients A, B, and C are derived empirically:

$$log_{10} P = A - \frac{B}{C + T} \quad (7)$$

There are a few candidate propellants examined for potential use (see Figure 7). Considerations include high chamber pressure with a low temperature and material compatibility with system components.

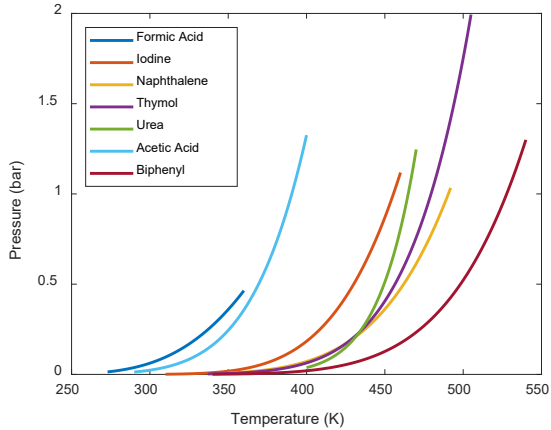

**Figure 7: Vapor Pressure vs. Temperature of Sublimate Substances (data from NIST [14])**

Sublimation is an endothermic process, requiring heat addition often by resistive heating for molecules to escape attractive forces. As the temperature of the system increases, the vapor pressure increases. This leads to one of the large advantages of using sublimating propellant as opposed to gas storage: the sublimate acts as its own pressure regulator through temperature control (until a sufficiently low mass of solid is left later in the propulsion system's life). In a gas storage system, thrust monotonically decreases as propellant is used due to lower chamber pressures after each thrusting maneuver. The sublimation vapor pressure is more stable, thus decreasing complexity of the system. There is no need for an external pressure regulator or for the system to be externally pressurized. Constant chamber pressure increases reliability of the system, resulting in lower error in delivered thrust.

The resultant propulsion system is shown in Figure 8. Storage density of a sublimate substance is much higher than gas storage, leading to a longer mission lifetime. The vapor pressure tends to be low compared to gaseous storage [15], up to a few bars for more volatile substances such as dry ice or urea. The positive effect of this is a simpler storage tank design, as the tank can have thinner walls and be lighter. Leak rate will also be slower, as that is dependent on the storage pressure. Lower chamber pressure does decrease the deliverable thrust however, as mass flow rate is proportional to chamber pressure. This will tend to limit sublimate systems to operation in a vacuum, such as attitude control or asteroid exploration.

There are two added design complexities of utilizing sublimate propellant. The storage tank must be temperature regulated and very well insulated to ensure proper control of chamber conditions. Heating of the tank can be done through Joule heating, such as nichrome wires wrapped around the tank or an incandescent film inserted into the propellant/ gas, such as in the Aerojet MR-501 rocket engine. Heating increases specific impulse, as specific impulse is proportional to chamber pressure.

To prevent two phase flow, as there will almost certainly be small particles of solid propellant flowing with the gas, a filter must be used upstream to prevent clogging of downstream components. Pressure losses induced by a filter will negatively affect thrust, so careful consideration of minimal impact must be balanced with filtering the required solid particulates. The mesh sizing will change dependent on each mission's requirements, largely the geometry and sizing of the propulsion system. The figure 6 is a representative schematic of the system, highlighting the few components and simplicity,

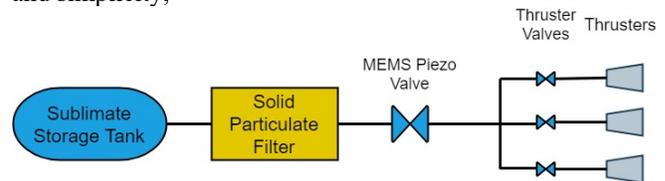

**Figure 8: Propulsion System Schematic**

## 5. GEOMETRIC APPROXIMATION OF NOZZLE

The design of the nozzle begins with isentropic flow relations [11]. The assumptions in derivation are homogeneous, perfect, single phase, adiabatic gas flow, a lack of discontinuities from shock waves, axial exhaust flow, and friction/ boundary layer effects are neglected. The working gas is nitrogen, as properties are well-known compared to sublimated gases.

The driving factor in determining thruster geometry is the Δv of each hop, with a maximum Δv sufficiently lower than the asteroid's escape velocity to ensure the robot is never lost to free space. A representative calculation for the asteroid, with equatorial radius of 400 m, and mass ~5*10¹¹ kg. Thus, the Δv is 0.129 m/s. A safety factor is applied to ensure escape velocity is not reached. This design thrust is the total thrust of all nozzles in the array, so the thrust of each nozzle is dependent on the number of nozzles. For calculation purposes, an array of 5 nozzles is examined. Using this corrected Δv, an assumed system mass of 1 kg (typical for 1U cubesats), and a burn time of 1 ms, the total thrust required can be calculated as

$$F = \frac{m \Delta v}{t_{burn}} \quad (8)$$

$$F = 0.1 \, mN$$

From this thrust result, equation (1) is used, assuming near optimal expansion so the pressure term may be neglected (in reality, the nozzle would be underexpanded due to vacuum ambient pressure on an asteroid). Thus, for a conical nozzle the thrust becomes

$$F = \lambda \dot{m} v_e$$

From Kundu et al. [16] the conical semi-vertex angle should be 28° as opposed to the normal optimal angle of 15° for macro-nozzles due to microfluidic effects. However, other angles are considered due to limitations in MEMS fabrication processes. Anisotropic etching results in pre-defined nozzle angles due to dependence on crystal structure of the substrate [17]. For example, silicon wafer anisotropic wet etching results in an angle of 35.3°, the angle of the <1 1 1> plane to vertical. A chamber pressure of 1 bar is used in further calculations, as it is easily attainable as shown in Figure 5. A design exit pressure of 100 Pa is used. Further correction due to viscous effects are needed, as the flow will have a low Reynolds number and a boundary layer displacement thickness approaching the order of the nozzle diameter [2]. These effects are captured in a corrected thrust coefficient $C_F$,

$$C_F = C_{F_i} - C_{F_v} \quad (9)$$

Where $C_{F_i}$ is the isentropic thrust coefficient,

$$C_{F_i} = \frac{F_i}{p_c A_t}$$

And $C_{F_v}$ is the thrust coefficient loss due to viscous effects. The thrust coefficient is defined as

$$C_F = \frac{F}{p_c A_t}$$

From experimentation on conical nozzles with cold gas flow [Spisz], the viscous loss term can be determined as

$$C_{F_v} = \frac{17.6 e^{.0032 \varepsilon}}{\sqrt{Re_{t,w}}} \quad (10)$$

Where $\varepsilon$ is the nozzle exit to throat area ratio and $Re_t$ is the Reynolds number at the throat near the wall.

$$Re_{t,w} = Re_t \left(\frac{T_t}{T_{t,w}}\right)^{\frac{5}{3}}$$

Where $Re_t$ is the Reynolds number at the throat, $T_t$ is the throat temperature, and $T_{t,w}$ is the throat wall temperature. The temperature ratio is assumed [18],

$$\left(\frac{T_t}{T_{t,w}}\right)^{5/3} = 0.857$$

Reynolds number at the throat, assuming ideal gas flow as an approximation,

$$Re_t = \frac{v_t d_t}{\nu_t}$$

Where $v_t$ is the throat flow speed (Mach = 1), $d_t$ is the throat diameter, and $\nu_t$ is the throat kinematic viscosity. Under ideal gas flow, the throat speed is sonic,

$$v_t = \sqrt{\gamma R T_t}$$

Kinematic viscosity, using the ideal gas law, can be defined as

$$\nu = \frac{\mu}{\frac{p}{RT}}$$

Kinematic viscosity at the throat is related to the stagnation condition through temperature,

$$\frac{\nu_t}{\nu_0} = \frac{\mu_t}{\mu_0} \frac{p_0}{p_t} \left(\frac{T_t}{T_0}\right)$$

From Spisz et al. [18], dynamic viscosity is related to temperature by

$$\frac{\mu_t}{\mu_0} = \left(\frac{T_t}{T_0}\right)^{\frac{2}{3}}$$

From isentropic relations, the pressure ratio can be converted to a temperature ratio

$$\frac{p_0}{p_t} = \left(\frac{T_0}{T_t}\right)^{\frac{\gamma}{\gamma-1}}$$

Thus, the kinematic viscosity ratio is

$$\frac{\nu_t}{\nu_0} = \left(\frac{T_t}{T_0}\right)^{\frac{-\gamma}{\gamma+1}+\frac{5}{3}} \quad (11)$$

From the stagnation to throat temperature ratio,

$$\frac{T_t}{T_0} = \frac{1}{1+\frac{\gamma-1}{2}M^2} = \frac{2}{\gamma+1}$$

Where at the throat, M = 1 (sonic). Thus, given some initial kinematic viscosity, the Reynolds number at the throat can be approximated by

$$Re_t = \sqrt{\gamma R T_t}\, d_t \left(\frac{T_t}{T_0}\right)^{\frac{-\gamma}{\gamma+1}+\frac{5}{3}} v_0^{-1}$$

The Reynolds number at the wall is found by

$$Re_{t,w} = 0.857\sqrt{\gamma R T_t}\, d_t \left(\frac{T_t}{T_0}\right)^{\frac{-\gamma}{\gamma+1}+\frac{5}{3}} v_0^{-1} \quad (12)$$

The area ratio is limited by MEMS manufacturing techniques; although large expansion ratios increase efficiency, they defy the purpose of miniaturizing nozzles and silicon wafers tend to be less than 775 microns thick. Stacking of wafers to increase nozzle length will inherently produce manufacturing errors due to wafer alignment limitations, so the thickness should be limited to a non-optimal state. Correction to the mass flow rate from isentropic conditions is done iteratively through the discharge coefficient $C_D$ [19],

$$\dot{m} = C_D \dot{m}_i \quad (13)$$

$$C_D = 0.8825 + 0.0079 \ln(Re_t)$$

With these correction factors for low Re flow, MATLAB is used to determine geometric properties of the divergent section of the converging-diverging nozzle. Analysis begins by calculating the desired thrust from equation (8) and fixing chamber pressure at 1 bar and chamber temperature at 400 K. Ranges of throat area and area ratio are used to calculate a range of exit areas

$$A_e = \varepsilon A_t \quad (27)$$

From this geometry, the expected viscous loss thrust coefficient is calculated based on the throat Reynolds number. Nitrogen is assumed stored at 400 K, 1 bar with kinematic viscosity $26.35 \times 10^{-6}$ m$^2$/s. Knowing the real thrust coefficient and loss coefficient, the isentropic thrust coefficient can be calculated.

$$C_{F_i} = C_F + C_{F_v}$$

The chamber pressure and throat area are the same for both the real and isentropic geometry. Thus, the required isentropic thrust required is calculated by

$$F_i = C_{F_i} p_c A_t = \lambda \dot{m} v_{e,i}$$

Mass flow rate is calculated by equation (13). Utilizing continuity,

$$\dot{m}_t = \dot{m}_e = \rho_e A_e v_e$$

Thus, for some exit area calculated from throat area and area ratio, the required exit velocity can be calculated as

$$v_{e,i} = \frac{\dot{m} R T_e}{p_e A_e}$$

A combination of isentropic nozzle geometry, exit velocity, and mass flow rate corresponds to some real thrust.

The geometry must produce the desired true thrust, shown in the dashed line on the thrust plot in Figure 7. A geometry is chosen to minimize the size while providing a large enough $I_{sp}$. For CFD analysis, a throat diameter of 700 microns with an expansion ratio of 7 is chosen to provide a thrust of 0.057 N. This results in an exit diameter of 4.9 millimeters per nozzle.

## 6. 2-D FLOW ANALYSIS

This geometric approximation is used to model a 2-D part in CAD with an additional converging portion of the nozzle. The converging section typically does not have a large effect on the fluid flow compared to the diverging section, though geometry will be investigated in later simulations. This model is then imported to ANSYS FLUENT for flow analysis. The working gas is assumed to be nitrogen, as well-known fluid models have been developed for nitrogen gas flow. Future FLUENT models will include sublimate gases, but experimentation is required to determine more fluid flow parameters such as specific heats and viscosity.

Turbulence modelling utilizes the k-ω two equation model, largely for its ability to predict flow separation and increased boundary layer considerations. A turbulence model is required due to Reynolds number on the order of 500-10,000.

The throat has a sharp turn, so it is possible for the flow to separate after the throat, which needs to be examined. Quadrilateral mesh elements are used for this relatively simple geometry, as shown in Figure 9. Mesh density is increased around the throat.

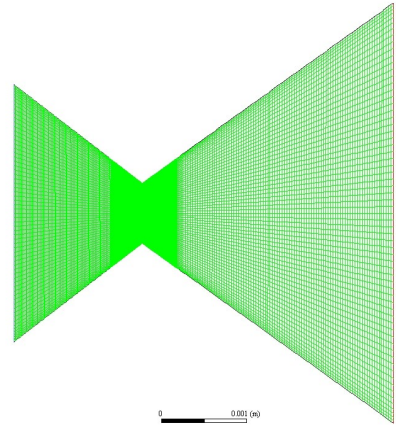

**Figure 9: Mesh Grid**

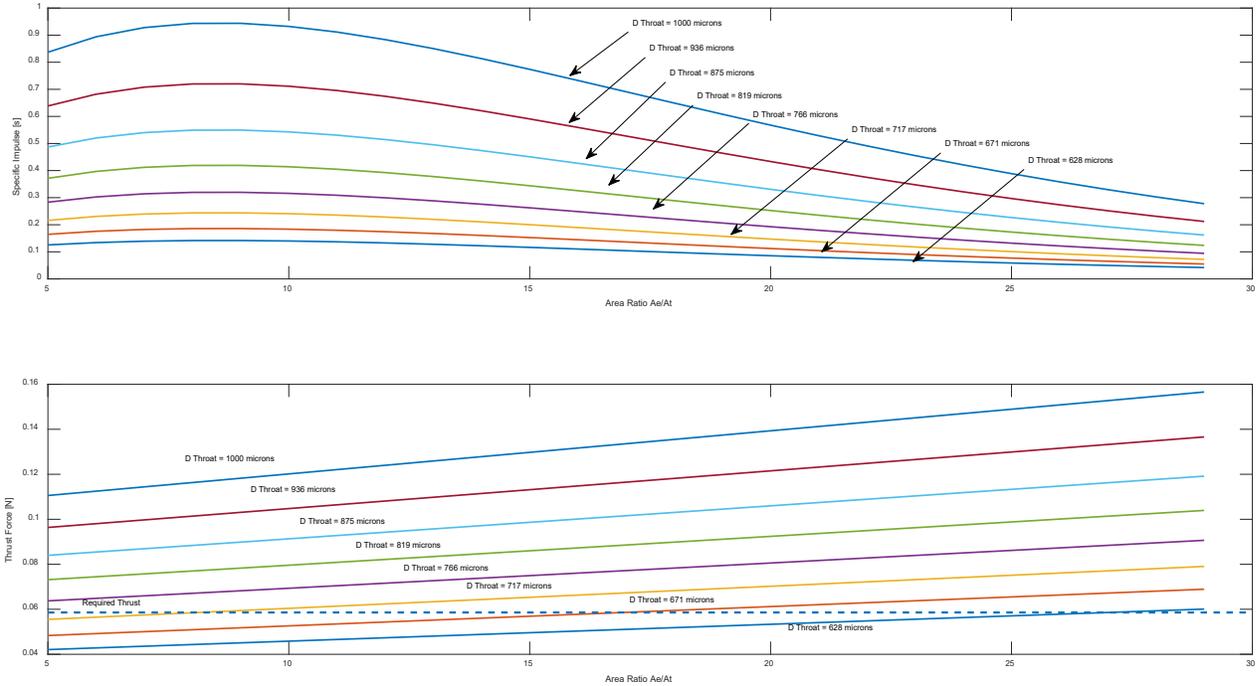

**Figure 10: $I_{sp}$ and Thrust for Various Area Ratios and Throat Diameters**

N₂ is used as the operating fluid. The operating pressure is set to 0 Pa. Table 3 shows the fluid properties and Table 4 shows the Physics models used.

**Table 3: Fluid Properties**

| Property | Value |
|---|---|
| Fluid | Nitrogen |
| P Inlet | 1 bar |
| T Inlet | 400 K |
| P Outlet | 100 Pa |
| Viscosity | Sutherland |
| $C_P$ | Piecewise |

**Table 4: Physics Models**

| Model | Use |
|---|---|
| Energy Eq. | On |
| Viscous Model | BSL k-ω |

All other models are turned off. Default values for the turbulent viscous model are used. For calculation, a Courant number of 0.2 is used. A design of experiments will be used to vary through different geometries. Figure 10 shows the Isp and thrust for various area ratios and throat diameters. Figure 11 shows the absolute pressure through the throat for a specific simulation.

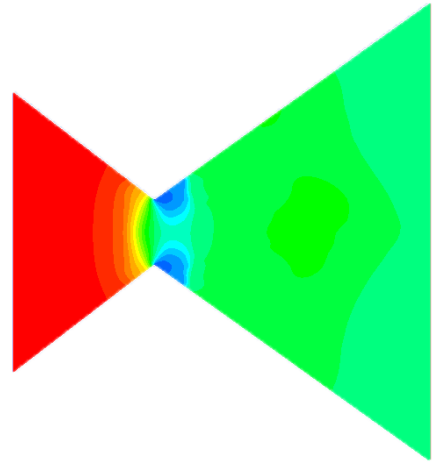

**Figure 11: Absolute Pressure**

The exit pressure is ~7500 Pa, three times higher than predicted. Updates to the numerical calculations must be updated iteratively to converge on a nozzle geometry to produce desired thrust. From the simulation results, there is flow separation after the throat of the nozzle. This indicates more complex geometries for the throat must be examined, particularly either changing the semi-vertex angle or adding a radius of curvature to the throat. Differing cone angles means other substrates must be examined, or other micromachining methods.

As previously mentioned, the length of the nozzle is a limiting factor for a selected design. Assuming the same wafer diameter that can fit each nozzle in the array despite throat and exit areas, the nozzle geometry with minimum length that still provides adequate and efficient thrust shall be selected. This reduces the mass of the nozzle array chip.

## 7. DISCUSSION

The fluid models for various sublimates must be improved to obtain accurate predictions of flow in FLUENT. This includes compiling information on specific heats and viscosity information, potentially through in-house experimentation. Prototypes of the nozzles will be constructed for testing of various sublimates.

This paper has only included design of the nozzle, not of ancillary components. These include the storage tank with temperature control and pressure sensors, a piezoelectrically actuated isolation valve, flow channel geometry, individual thruster actuation valves, and interfaces between macro and micro devices. A piezoelectric valve will be used because of the high sealing forces that can be obtained, reducing loss of propellant during transit to an asteroid.

To ascertain control authority of the propulsion system on AMIGO for attitude control during a hop, the dynamics of AMIGO must be further characterized. AMIGO entails three mobility modes after landing on an asteroid in which propulsion may be useful: ballistic hopping, attitude control during a hop, and uprighting maneuvers after landing potentially not upright. Most of the mass and placement of the hardware is known, from sources of commercial off the shelf solutions and science instruments to be developed. However, the inflatable spheroid will contribute significant inertia and variability due to its size and potential deformability. This deformability can potentially cause instabilities along the intermediate and minor inertia principle axes.

## 8. CONCLUSIONS

In this paper, nozzle geometry for an array of sublimate-based MEMS thruster system with corrections for low thrust, low Reynolds number flow is designed for representative parameters. Beginning with a known required thrust based on max Δv requirements, a nozzle is designed to minimize the length while still providing enough thrust efficiently.

Generally, the $I_{sp}$ of sublimate propellants will be much lower than other cold gas storage schemes. However, the density $I_{sp}$, a measure of the efficiency of some volume of stored propellant, will be of the same order due to the much higher storage densities. For example, the density at standard temperature and pressure of urea is $1.32 \times 10^3$ g/L, while the density of nitrogen is 1.25 g/L. The density $I_{sp}$ is a better measure for overall system performance taking into account the tank properties. For a constant volume of propellant between gas storage and sublimates, the density $I_{sp}$ will be on the same order. When accounting for the lower mass of the tank due to lower experience pressures, a complete sublimate system will save weight over other storage methods.

## ACKNOWLEDGEMENTS

The authors would like to thank Dr. Stephen Schwartz at the Lunar and Planetary Laboratory (LPL) for assistance in developing the AMIGO concept.

**BIOGRAPHY**

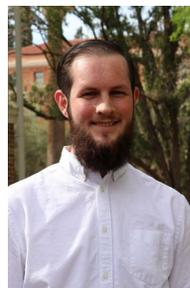

*Greg Wilburn* received his B.S. in Aerospace Engineering from the University of Arizona in 2018 and is currently a Master's student. He has worked for the Space and Terrestrial Robotics Exploration Lab (Space TREx) at the University of Arizona since 2017 as a research assistant. Greg has worked on multiple systems for proposed cube sat missions, including power budget analysis and small propulsion devices. His main research is in miniaturizing propulsion systems to enable more complex small satellite missions.

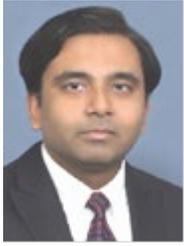 *Jekanthan Thangavelautham* has a background in aerospace engineering from the University of Toronto. He worked on Canadarm, Canadarm 2 and the DARPA Orbital Express missions at MDA Space Missions. Jekan obtained his Ph.D. in space robotics at the University of Toronto Institute for Aerospace Studies (UTIAS) and did his postdoctoral training at MIT's Field and Space Robotics Laboratory (FSRL). Jekan is an assistant professor and heads the Space and Terrestrial Robotic Exploration (SpaceTREx) Laboratory at the University of Arizona. He is the Engineering Principal Investigator on the AOSAT I CubeSat Centrifuge mission and is a Co-Investigator on SWIMSat.